# Electron Bunching in Transport Through Quantum Dots in High Magnetic Field

O. Zarchin[1],[*] Y. C. Chung[2], M. Heiblum[1], D. Rorhlich[1] and V. Umansky[1]

[1] *Braun Center for Submicron Research, Dept. of Condensed Matter Physics,*

*Weizmann Institute of Science, Rehovot 76100, Israel*

[2] *Department of Physics, Pusan National University,*

*Buson 609-735, Republic of Korea*

## ABSTRACT

Shot noise measurements provide information on particles' charge and their correlations. We report on shot noise measurements in a ubiquitous quantum dot under a quantized magnetic field. The measured noise at the peaks of a sequence of conductance resonances was some nine times higher than expected; suggesting *bunching* of electrons as they traverse through the dot. This enhancement might be mediated by an additional weakly coupled level to the leads or an excited state. Note that in the absence of magnetic filed no bunching had been observed.

The 'constant interaction' model of quantum dots (QDs) lumps electron interactions in with a capacitance [1]. Consequently, due to the charging energy of the dot an electron will enter the dot only after another one leaves. This will lead to an 'ordered' stream with a suppressed shot noise [2]. Here, we report on shot noise measurements in QDs, under a high magnetic field (filling factor 1 and 2), with a highly enhanced shot noise. An enhancement of up to a factor of nine was found. This suggests transport of electrons in groups (bunched) through the dot. We propose that the unexpected bunching is mediated two-level transport.

Shot noise measurements in QDs are scarce [3-6], however, indications of bunching had been already reported. Multiple electron occupation of QDs was suggested via capacitance measurements [7]; enhanced shot noise was found in the negative differential resistance regime in resonant tunneling diodes [8]; and large shot noise was measured in a tunnel barrier, attributed to a nearly degenerate localized state [9]. In the latter case the proposed model invoked two participating levels, one of them coupled weakly to the leads, turning the current off when occupied, leading to supper-Poissonian noise and depending quadratically on the current.

At zero temperature, the quantum shot noise of a partitioned single channel in a ballistic conductor is [10,11]:

$$S_{T=0} = 2eV_{SD} g_Q t(1-t) \quad , \tag{1}$$

where $S$ the low frequency ($f<<eV_{SD}/h$) spectral density of current fluctuations ($S\Delta f=<i^2>$), $V_{SD}$ the applied source-drain voltage, $g_Q=e^2/h$ the quantum conductance, and $t$ the transmission coefficient of the channel. This reduces to the *classical* Poissonian expression when $t<<1$ ('Schottky equation'), $S_{Schottky}=2eI$, with $I=V_{SD}g_Qt$ the transmitted current. For a QD in the Coulomb blockade regime, shot noise may be either suppressed or enhanced. When $eV_{SD}>>\Gamma$ ($\Gamma$- energy width of the resonance), shot noise is suppressed [2, 12], with suppression approaching a factor of 2 in a symmetric device. Alternatively, for an elastic-cotunnelling process (QD remains in the same state), shot noise obeys Eq. (1), while for an inelastic-cotunnelling process (QD is left in an excited state), shot noise could be enhanced by up to a factor three [12, 14].

Since the conductance of a QD is not constant with applied voltage, thermal noise and back-injected amplifier's noise, which depend on the conductance, may dominate the noise [15]. Hence, applying a quantizing magnetic field, with edge channels transport, allows the amplifier (in a *multi terminal* configuration, see Fig. 1) to 'feel' a constant conductance, $\nu e^2/h$, with $\nu$ the filling factor. Consequently, the spins are polarized and the single particle spectrum in the dot turns into magneto-electric spectrum [16-18], with a new configuration of charge distribution [19, 20].

Measurements were conducted with three different QD configurations on two different heterostructures (dots' size ~0.2x0.2μm$^2$), with high mobility 2D electron gases (density ~3·10$^{11}$ cm$^{-2}$, low temperature mobility ~8·10$^5$ cm$^2$/V·s). The QDs were formed by biasing metallic gates deposited on the surface of the heterostructures (see Fig. 1). The different configurations behaved in a similar manner for $\nu$=1 & 2 in the bulk; here we present a detailed account of our studies with one configuration, at electron temperature $T$~25mK (determined from measured $4k_BTg$, see Ref. [21]) and $\nu$=1 ($B$=9-11 Tesla). Conductance was measured by two methods: (1) applying 0.5μV excitation voltage at ~0.9 MHz and amplifying the output signal with a cold (in situ) amplifier (with 75kHz bandwidth determined by a cold LC circuit); (2) applying 2μV excitation voltage at 7Hz and employing a standard Lock-In technique. The noise was measured at ~0.9MHz for a DC voltage applied at the source (see Refs. [21, 22] for details).

In Fig. 1(a) we show a sequence of conductance peaks, reaching almost the conductance limit $e^2/h$, as function of plunger gate voltage $V_P$. The plunger gate voltage pinches the source and drain quantum point contacts (QPCs) via electrostatic induction, hence the affecting the symmetry of the dot and the peak heights. Its direct effect on dot's potential was determined by measuring the conductance as function of $V_P$ and $V_{SD}$ (constructing a 'diamond structure' [23]; Fig. 1(b)). We extracted a *levering factor* ~16 (the plunger gate voltage in units of dot's potential), charging energy $U_C$~500μeV, and single particle level spacing $E_{ls}$~100μeV.

We measured shot noise as function of $V_{SD}$, however, not like in previous works [3-6], $V_{SD}$ was smaller than the level width (15-100μeV). In Fig. 2 we provided plots $S$ versus $V_{SD}$ measured across the full width of a typical conductance peak. As $V_{SD}$ exceeded ~$2k_BT$ shot noise grew and approached an almost linear dependence on $V_{SD}$. Very surprisingly, at the conductance peak the magnitude of the noise was much higher than expected. Away from the peak the noise got smaller, approaching the expected value at the tails of the peak.

What is the expected value of the noise? We first modify Eq. (1) to allow for finite temperature [24]:

$$S(V_{SD},T) = 4k_BT g_Q + 2eV g_Q t(1-t)[\coth\chi - 1/\chi] , \qquad (2)$$

with $\chi = eV_{SD}/2k_BT$. However, when the transmission is bias dependent, the shot noise $S_{excess} = S(V_{SD},T) - 4k_BT g_Q$ for a certain $V_{SD}$, can be approximated by $S_{excess} = \int_{V_{SD}=0}^{V_{SD}} dS$, where $dS = S(V_{SD}+dV,T) - S(V_{SD},T)$ and the differential transmission $t(V_{SD})$ is approximated by $t(V_{SD}) = [I(V_{SD}+dV) - I(V_{SD})]/(g_Q dV)$ - in each $dV$ interval $t(V_{SD})$ can be regarded constant.

While in general the Fano factor $F$ is defined as the ratio between the measured noise $S'_M=S_M(V_{SD})-S_M(0)$ and the classical shot noise $F= S'_M /2eI$ [3, 4], we defined a modified Fano factor that relates to the expected shot noise:

$$F^*=S'_M /S_{excess} , \qquad (3)$$

where $F=F^*(1-t)$. At zero temperature we can define $e^*=F^*e$, and interpret $e^*$ as an effective charge that is being stochastically partitioned; namely, $e^*/e$ electrons are being bunched in the transport process.

Returning to Fig. 2 we note that the transmission at the peak of the resonance is $t=0.75$ and $F^*\sim9$ (or $F\sim2$); namely, the current is carried by bunches of approximately ~9 electrons. Away from the peak maximum, toward the conductance valley, $F^*$ drops, reaching $F^*\sim1$ (and $F\sim1$) in the tails. Increasing the width of the conductance peak from 15μeV to 100μeV (by opening the constrictions of source and drain), does not affect in any significant way $F^*$ across the peak.

However, this is not the case in an asymmetric QD. Figure 3 summarizes the dependence of the Fano factor $F^*$ on the deviation of the QD from symmetry, with similar results obtained when keeping the opening the drain (source) constriction constant and varying that of the source (drain). The ratio $\Gamma_d/\Gamma_s$, where $\Gamma_s$ ($\Gamma_d$) denotes the coupling (in units of energy) to the source (drain) lead, was estimated via [2]:

$$\frac{g}{g_Q} = 4\frac{\Gamma_s\Gamma_d}{(\Gamma_s + \Gamma_d)^2} . \qquad (4)$$

As the dot was tuned away from symmetry, the conductance peak as well as the shot noise quenched, eventually reaching $F^*\sim1$ throughout the conductance peak.

An important clue is that we never were able to reach the conductance limit at the peak of $e^2/h$. This was despite the fact that the electron temperature was much smaller than the level width. This fact may suggest the following possibilities: (a) A strong fluctuating potential turning the dot on and off, hence lowered the average current; (b) A weak fluctuating potential, on the scale of the level width, leading to an averaging of the transmission and lowered the peak current; (c) An additional energy level that is weakly coupled to the leads (hence, a long life time), partly degenerate with the main, strongly coupled, level to the leads. When this level is occupied, it blocks the transport through the main level due to Coulomb blockade [9]; and (d) An excited state with a tail partly overlapping the main level, which is populated with a lower probability, with a resultant effect to that in (c).

A strong fluctuating potential can mimic bunching, however, the noise will then depend quadratically on the applied voltage (or the impinging current); not according to our measurements (see Fig. 2b). A weak fluctuating potential will lead to noise that depends on the derivative of the differential conductance with respect to the dot's potential (or to the plunger gate voltage). The noise should then be close to zero at the conductance peak maximum and high somewhere along the peak side. This also contradicts the observed behavior, with the highest Fano factor at the peak's maximum. On the other hand, the *two level* model (be it a localized state or an excited state) may offer a better explanation. Assume a probability $P_{in}$ to enter the localized level, thus transport through the main level will be blocked after an average number of $1/P_{in}$ electrons impinge at the dot (during time $h/eV_{SD}P_{in}$), and electrons transport will be bunched to an average charge $e^*=e/P_{in}$. Since the number of electrons in a single bunch depend on the applied voltage, shot noise will depend linearly on the applied voltage with charge $e^*$ replacing $e$. A similar effect was suggested to explain an enhanced noise in a tunnel barrier containing an impurity, however there, the applied voltage was larger than the resonance width and the noise depended quadratically on the current [9].

Is the enhanced noise a direct result of the applied magnetic field? Removing the magnetic field necessitates a 'two terminal' configuration (injecting from source and measuring shot noise in the drain, see caption of Fig. 1) [21]; with a conductance dependent thermal noise voltage ($4k_BT/g$) and an amplifier's 'back injected' noise (($i_{amp})^2/g^2$) - both are added to the sought after shot noise. Since the latter contributions are of the same order of magnitude as that of the shot noise, our measurement carried much larger error bars than the measurements with magnetic field. However, we concluded that the shot noise, at zero magnetic field and up to that leading to filling factor 4 was the expected one, namely, $F^*\sim1$! As the magnetic field was further increased, the Fano factor grew, approaching $F^*\sim9$ at $\nu=2$. These measurements further confirm that external fluctuations are not responsible for the large Fano factor (otherwise, enhancement would be observed also at $B=0$).

Can this observation shed light on the mechanism that leads to bunching? Since the number of electrons in our dot is ~120, the average level spacing is roughly 80µeV. As the magnetic field increases the confined states are modified (to magneto-electric states), they approach and may even cross each other, eventually merging into Landau levels [16-18]. Moreover, a rearrangement of the charge leads to a self-consistent distribution with a likely maximum density droplet in the center of the dot (at small filling factors [19, 20]). Either overlapping states (due to the smaller spacing) or an electron droplet weakly coupled to the leads (hence, a localized state) could serve as the second, long lived, level we are looking for.

In Fig. 4 we plotted, in a color scale, the measured noise as a function of $V_{SD}$ and $V_P$. For comparison, we also plotted the expected shot noise $S_{excess}$ (being smaller than the measured noise, its scale is smaller by a factor of 6). The measured noise peeked more sharply, as consequence of the increasing $F^*$ at the conductance peak.

Our message here is that the shot noise, generated by a quantum dot in the Coulomb blockade regime and under a quantizing magnetic field (filling factor 1 or 2), is strongly enhanced. Hence, not like in the absence of magnetic field, where electrons are stochastically transferred one-by-one, bunching of electrons must take place as they traverse the dot. This surprising result suggests that transport through the QD is not via a single level. Another state, being more localized and nearly degenerate with the main state, captures the electrons infrequently and blocks the transport through the dot, hence, allowing bunches of electrons to pass between 'long' periods of blockaded transport.

**Acknowledgement**

We thank I. Neder, Y. Oreg, S. Gorvitz, E. Sela and M. Zaffalon for helpful discussions. O. Zarchin acknowledges support from the Israeli Ministry of Science and Technology. This work was partly supported by the MINERVA foundation, the German Israeli foundation (GIF), the German Israeli project cooperation (DIP), the Israeli Science foundation (ISF) and the Israeli - Korean program (P.N. M6050400004005B040004010).

**References**

* E-mail: oren.zarchin@weizmann.ac.il

[1] T. Ihn, Electronic Quantum Transport in Mesoscopic Semiconductor Structures (Springer-Verlag, 2004), Ch. 11.2.

[2] J. H. Davies, P. Hyldgaard, S. Hershfield, and J. W.Wilkins, Phys. Rev. B 46, 9620 (1992).

[3] H. Birk, et al., Phys. Rev. Lett. **75**, 1610 (1995).

[4] S. Gustavsson, et al., Phys. Rev. Lett. **96**, 076605 (2006).

[5] Yuan P. Li et al., Phys. Rev. **B 41**, 8388 (1990).

[6] E. Onac, et al,. Phys. Rev. Lett. **96**, 026803 (2006).

[7] N. B. Zhitenev, et al., Phys. Rev. Lett. **79**, 2308 (1997).

[8] G. Iannaccone, et al., Phys. Rev. Lett. **80**, 1054 (1998).

[9] S. S. Safonov et al., Phys. Rev. Lett. **91**, 136801 (2003).

[10] Y. M. Blanter and M. B¨uttiker, Phys. Rep. **336**, 1 (2000).

[11] G. B. Lesovik, Pis'ma Zh. Eksp. Teor. Fiz. **49**, 513 (1989) JETP Lett. **49**, 592 (1989).

[12] E. V. Sukhorukov, et al., Phys. Rev. **B 63**, 125315 (2001).

[13] S. De Franceschi et al., Phys. Rev. Lett. **86**, 878 (2001).

[14] A. Thielmann et al. Phys. Rev. Lett. **95**, 146806 (2005).

[15] J. B. Johnson, Phys. Rev. **29**, 367 (1927); H. Nyquist, Phys. Rev. **32**, 229 (1928).

[16] V. Fock, Z. Phys. 47, 446 (1928); C. G. Darwin, Proc. Cambridge Philos. Soc. **27**, 86 (1930).

[17] L. P. Kouwenhoven et al., Science **278**, 1788 (1997).

[18] N. C. van der Vaart et al., Phys. Rev. **B 55**, 9746 (1997).

[19] P. L. McEuen et al., Phys. Rev. **B 45**, 11419 (1992).

[20] P. L. McEuen et al., Phys. Rev. Lett. **66**, 1926 (1991).

[21] R. de-Picciotto et al., Nature **389**, 162 (1997) ; M. Reznikov et al., Nature **399**, 238 (1999).

[22] E. Comforti et al., Nature **416**, 515 (2002) ; Y. C. Chung, M. Heiblum and V. Umnasky, Phys. Rev. Lett. **91**, 216804 (2003) ; Y. C. Chung et al., Phys. Rev. B **67**, 201104(R) (2003).

[23] L. P. Kouwenhoven, D. G. Austing and S. Tarucha, Rep. Prog. Phys. **64**, 701 (2001).

[24] Th. Martin and R. Landauer, Phys. Rev. **B 45**, 1742 (1992).

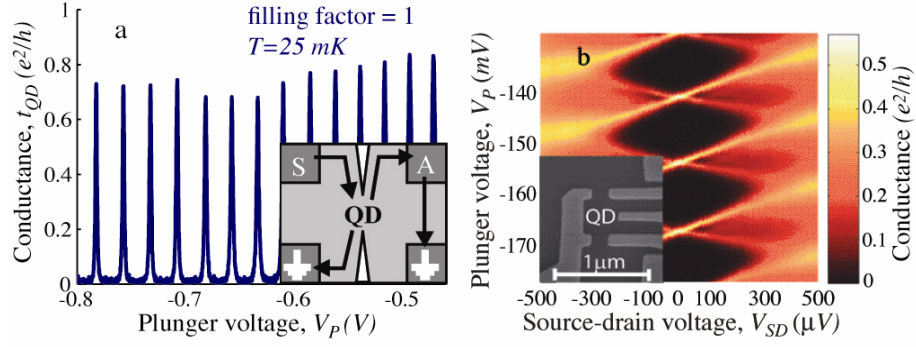

FIG. 1: Conductance of the quantum dot (QD). (a) Differential conductance of Coulomb blockade peaks as a function of plunger gate voltage ($V_p$), and (b) Diamond structure – differential conductance as a function of plunger gate voltage and source drain voltage ($V_{SD}$). Inset (a) – *Multi terminal* measurement scheme. Noise generated by the QD is measured at *A*. Due to edge state chirality, conductance from *A* to ground is kept constant; independent on the conductance of the QD [22]. For noise measurements at zero magnetic field, grounds are disconnected from the lower ohmic contacts, leaving a *two terminal* configuration, with an input and output conductances depending on the setting of the QD. Inset (b) – SEM picture of the QD.

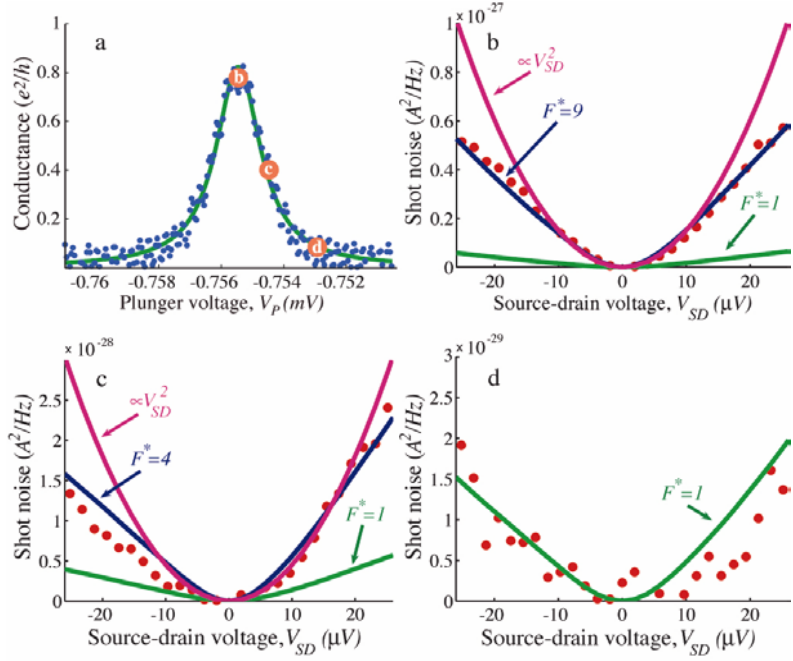

FIG. 2: Conductance and shot noise. (a) Differential conductance (blue circles) of a single resonance peak, and (b-d) shot noise as function of source-drain voltage ($V_{SD}$) for three different plunger gate voltages, with $F^*$ represents the ratio between the measured shot noise $S'_M$, to the theoretical shot noise $S_{excess}$. In (a), green line represents a lorentzian fit while orange circles represent the location where shot noise was measured in b-d. In b-d the blue lines represent fits to the measured shot noise (red circles) with $F^*=9$ (b), $F^*=4$ (c), $F^*=1$ (d). The green line is a theoretical fit with $F^*=1$. The red line (in a, b), $S \propto V_{SD}^2$, represents the effect of classical fluctuations.

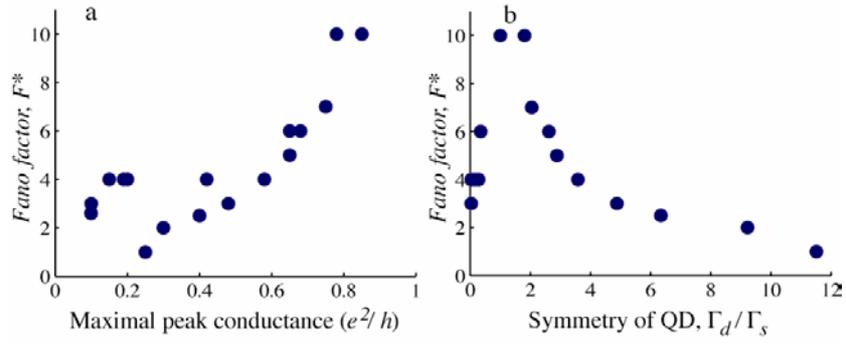

FIG. 3: Enhancement of shot noise. (a) $F^*$ as function of maximal peak conductance (via varying the QPC on the drain side). (b) Dependence of $F^*$ on $\Gamma_d/\Gamma_s$, where $\Gamma_s$ ($\Gamma_d$) is the coupling to the source lead (drain lead), estimated via Eq. (4). For a highly symmetric peak $F^*\sim 10$; dropping to $F^*\sim 6$ at $\Gamma_d/\Gamma_s\sim 0.3$ or $\sim 3$, and toward $F^*\sim 1$ as the asymmetry grows.

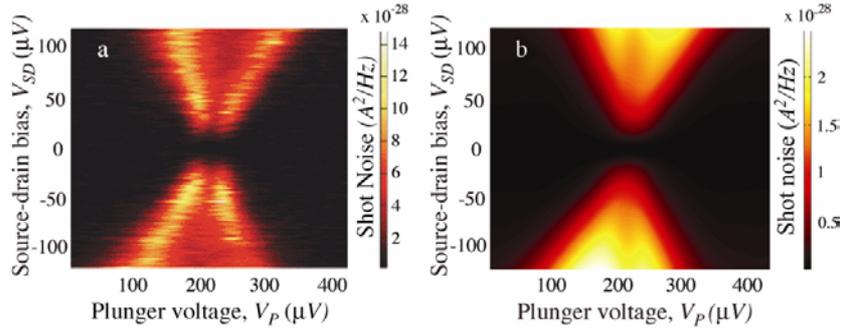

FIG. 4: Shot noise dependence on source-drain bias ($V_{SD}$) and plunger gate voltage ($V_P$). (a) Measured Shot noise $S'_M$, and (b) Calculated shot noise $S_{excess}$. The shot noise scale in (b) is 6 times smaller than in (a) indicating a much larger shot noise than expected for $F^*=1$. The QD transmission behaves similarly to that in Fig. 1b.